\documentclass[twocolumn]{aastex631}

\usepackage{amsmath}
\usepackage{graphicx}
\usepackage{url}
\usepackage{natbib}
\shorttitle{Independent Transient Detection on POSS-I Plates}
\shortauthors{Hayes}

\begin{document}

\title{Independent Recovery of Vanishing Sources on POSS-I Photographic Plates
Using Automated Source Detection and Cross-Epoch Matching}

\author{Zachary Hayes}
\affiliation{Independent Researcher}

\begin{abstract}
We present an independent pipeline for detecting candidate vanished
sources on digitized first-epoch Palomar Observatory Sky Survey
(POSS-I) photographic plates. The pipeline detects and PSF-filters
sources on POSS-I Red DSS cutouts, applies local astrometric
registration refinement, and identifies candidates by cross-epoch
matching against POSS-I Blue and POSS-II Red with Pan-STARRS~DR1
rejection. On a 20-case benchmark harness, the pipeline recovers 8/9
sources in the April 1950 field and 3/3 in the July 1952 field, with a
false positive rate of ${\sim}0.2$ per 10~arcmin field on random
non-crowded controls. A full-footprint sweep over the POSS-I coverage using
30~arcmin patches yields a filtered catalog of 2.85~million candidate
vanished sources after post-processing PSF cuts, deduplication, and
Pan-STARRS~DR1 rejection. Cross-matching against the 5{,}399-source
\citet{solano2022} catalog yields 3{,}450 matches (63.9\%) with median
separation 0.94~arcsec; among unrecovered catalog entries within our
footprint, we find no Pan-STARRS~DR1 counterpart within 3~arcsec.
Applying \citet{bruehl2025}-style temporal windows to this catalog over
the 368 POSS-I observation nights in the 1949--1957 interval gives a
post-test calendar-day relative risk of 1.35 for the $[+1]$ window,
but the effect is not statistically significant (95\% CI [0.91, 2.00];
two-sided Fisher $p = 0.17$) and is sensitive to coding unobserved days
as zero-transient days. A negative binomial model of nightly candidate
counts with nightly patch coverage as exposure is likewise null
(IRR~$= 1.03$, 95\% CI [0.89, 1.18], $p = 0.71$). The catalog-level
replication is strong; the temporal association remains inconclusive.
\end{abstract}


\section{Introduction}
\label{sec:intro}

The Vanishing and Appearing Sources during a Century of Observations
(VASCO) project has cataloged thousands of candidate transient sources
by comparing historical sky survey plates against modern digital
archives \citep{villarroel2020}. Among the most striking results are
nine simultaneously appearing and vanishing point sources on a single
POSS-I Red plate from April 1950 \citep{villarroel2021}, a bright
triple transient from July 1952 \citep{solano2024}, and a
publicly available catalog of 5{,}399 vanishing-source candidates
identified on POSS-I plates \citep{solano2022}.

The nature of these transients remains debated: while some authors
have attributed the narrower, rounder profiles of the transient
candidates to emulsion flaws, \citet{villarroel2025profiles} argue
that sub-second optical flashes naturally produce sharper, more
circular profiles than stars on long-exposure plates due to reduced
seeing and tracking blur. Meanwhile, \citet{bruehl2025} reported a
statistical association between the temporal distribution of POSS-I
transient candidates and above-ground nuclear weapons tests conducted
during the 1949--1957 survey period. Their analysis found an elevated transient rate within
$\pm$1~day of nuclear test dates, raising the question of whether
some fraction of photographic-plate transients may be attributable
to environmental contamination rather than astrophysical phenomena.

Independent replication of these results using different detection
methods is essential for assessing their robustness. In this work, we
describe a fully automated pipeline that detects sources on digitized
POSS-I plates, filters by point-spread function (PSF) morphology,
performs local astrometric refinement, and identifies candidate
vanished sources via cross-epoch matching---without reference to
existing transient catalogs. We validate the pipeline against known
benchmark fields and cross-match it against the \citet{solano2022}
catalog to quantify independent recovery rates.

\section{Method}
\label{sec:method}

\subsection{Data}

We use digitized plate cutouts from the STScI Digitized Sky Survey
(DSS) archive \citep{lasker2008}, with embedded WCS headers for three
epochs: POSS-I Red ($\sim$1949--1957), POSS-I Blue (same epoch), and
POSS-II Red ($\sim$1985--2000). Benchmark cases are evaluated at their
case-specific cutout sizes (4--10~arcmin). The production sweep uses a
30~arcmin patch grid with $\cos(\delta)$-corrected spacing over the
full POSS-I footprint (Dec~$> -33\degr$), yielding 122{,}991 patch
positions across 932 plates.

\subsection{Source Detection and Filtering}

For each cutout, sources are detected via sigma-clipped thresholding
above the local background, followed by connected-component labeling.
Each detection is characterized by its full width at half maximum
(FWHM) and ellipticity. A PSF morphology filter retains only sources
within a specified FWHM range relative to the plate median and below
a maximum ellipticity, rejecting extended objects, plate artifacts,
and noise peaks.

The detection threshold and filter bounds were calibrated on a 20-case
benchmark harness (Section~\ref{sec:benchmark}). The resulting
calibrated settings ($8\sigma$, FWHM $0.7$--$1.5\times$ median,
ellipticity~$< 0.3$) reduce false positives to ${\sim}0.2$ per
10~arcmin field on random non-crowded control fields. The full-sky sweep
(Section~\ref{sec:fullsky}) was executed with an earlier set of
preliminary parameters ($5\sigma$, broader PSF acceptance) prior to
this calibration; the differences in catalog composition are discussed
there.

\subsection{Local Registration Refinement}

Although the DSS WCS headers provide global astrometric solutions,
spatially correlated residuals of 1--3~arcsec are common across
Schmidt plates. We compute a local affine correction for each
cutout by matching the 20--50 brightest isolated sources between
epochs and fitting a 6-parameter transformation with iterative
sigma-clipping. Post-correction median residuals are
0.2--0.5~arcsec, with 95th-percentile residuals below 1.3~arcsec.

\subsection{Transient Identification}

A source detected on POSS-I Red is flagged as a transient candidate
if no counterpart is found within 8~arcsec (after local registration
correction) on \emph{both} POSS-I Blue and POSS-II Red. The
requirement for absence in two independent reference epochs reduces
contamination from color-dependent detection limits and
epoch-specific artifacts. Given 95th-percentile post-correction
residuals below 1.3~arcsec, this radius is intentionally
conservative: it favors missed recoveries in crowded fields over
falsely labeling obvious cross-epoch counterparts as vanished.

\subsection{Pan-STARRS Verification}

For full-sky analysis, we additionally cross-match candidate
transients against Pan-STARRS~DR1 \citep{chambers2016} via CDS~XMatch.
Sources with a Pan-STARRS counterpart within 3~arcsec are rejected
as persistent objects that failed cross-epoch matching due to
astrometric or morphological differences between plate and CCD
data. This step removes obvious persistent objects and leaves a
catalog of candidate archive-absence sources. Because this screen
relies on a single modern survey and simple positional matching, we
refer to the retained sample as candidate vanished sources rather
than confirmed astrophysical disappearances.

\subsection{Synthetic Injection Testing}

To assess completeness and purity independently of historical
benchmark cases, we inject synthetic sources into plate cutouts.
Artificial stars are generated by scaling template PSFs from confirmed
real stars on each plate; artificial artifacts are generated as
sharp-edged Gaussian profiles lacking atmospheric wings. Across 120
star-injection trials, the pipeline recovers 79.2\% of injected
stars. Across 120 artifact-injection trials, 96.7\% of artifacts are
correctly rejected by the PSF filter. The star-recovery rate indicates
that the pipeline is useful for independent validation but not
exhaustive for population-level completeness claims.

\section{Results}
\label{sec:results}

\subsection{Benchmark Validation}
\label{sec:benchmark}

Table~\ref{tab:benchmark} summarizes results on a 20-case benchmark
harness. The pipeline recovers 8 of 9 known transients on the
April 1950 field \citep{villarroel2021} and all 3 transients on the
July 1952 field \citep{solano2024}. Blue-plate controls of both fields
return zero candidates, confirming that detections are
epoch-dependent. Of 6 random control fields at diverse Galactic
latitudes, 5 return zero candidates and 1 returns a single candidate.
Crowded Galactic plane fields show elevated false positive rates
(up to 22 per field), as expected from source confusion in
dense stellar environments.

\begin{deluxetable}{lcccc}
\tablecaption{Benchmark results with calibrated pipeline parameters
($\sigma = 8$, match radius $= 8''$, FWHM $= 0.7$--$1.5 \times$~median,
ellipticity $< 0.3$). \label{tab:benchmark}}
\tablehead{
\colhead{Field type} & \colhead{$N$} & \colhead{Expected} &
\colhead{Observed} & \colhead{FP rate}
}
\startdata
April 1950 Red    & 1 & 9 & 8 & --- \\
July 1952 Red     & 1 & 3 & 3 & --- \\
Blue-plate neg.   & 2 & 0 & 0 & 0.0 \\
Same-plate ctrl   & 8 & 0 & 0--10 & 2.0 \\
Random ctrl       & 6 & 0 & 0--1 & 0.2 \\
Crowded galactic  & 2 & 0 & 1--22 & 11.5 \\
\enddata
\end{deluxetable}

\begin{figure}[t]
\centering
\includegraphics[width=\columnwidth]{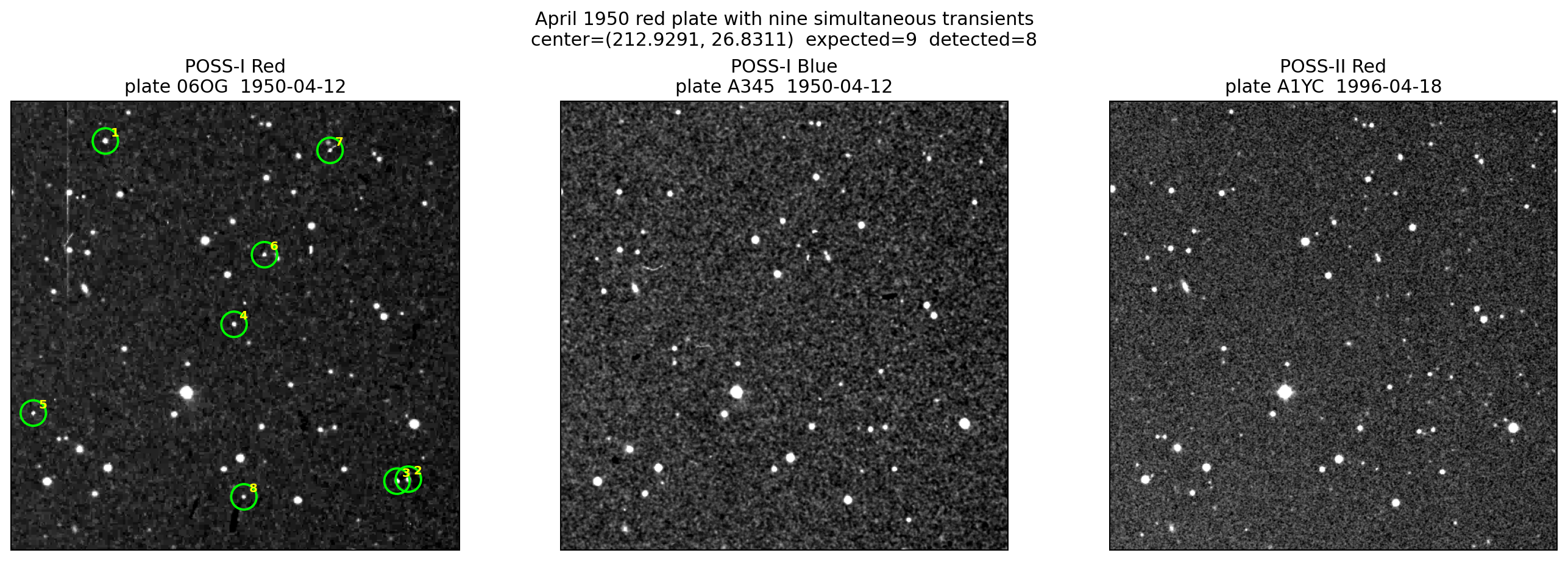}
\caption{April 1950 benchmark field. Left: POSS-I Red with 8 detected
transient candidates circled (of 9 known). Center: POSS-I Blue (same
date)---no candidates detected. Right: POSS-II Red ($\sim$1996)---no
counterparts at candidate positions. \label{fig:apr1950}}
\end{figure}

\begin{figure}[t]
\centering
\includegraphics[width=\columnwidth]{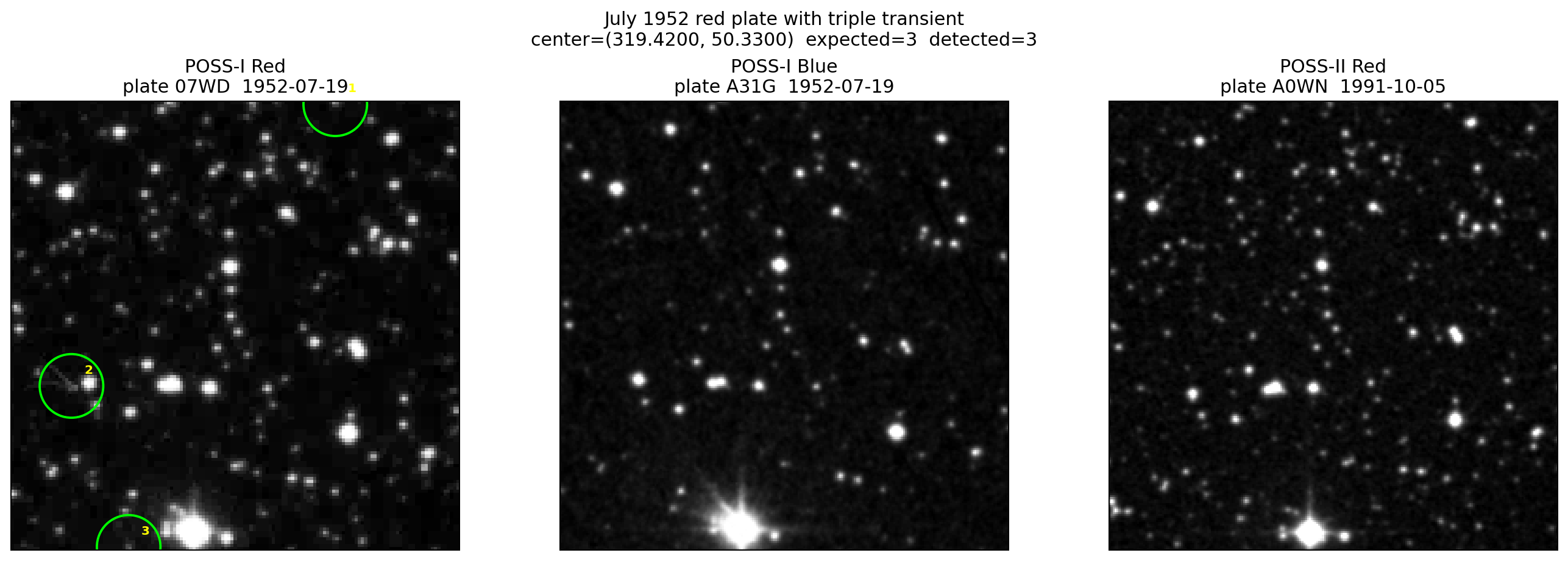}
\caption{July 1952 benchmark field. All 3 known transients from
\citet{solano2024} recovered. Same panel layout as
Figure~\ref{fig:apr1950}. \label{fig:jul1952}}
\end{figure}

\subsection{Solano Catalog Cross-Match}
\label{sec:solano}

\begin{figure}
\centering
\includegraphics[width=\columnwidth]{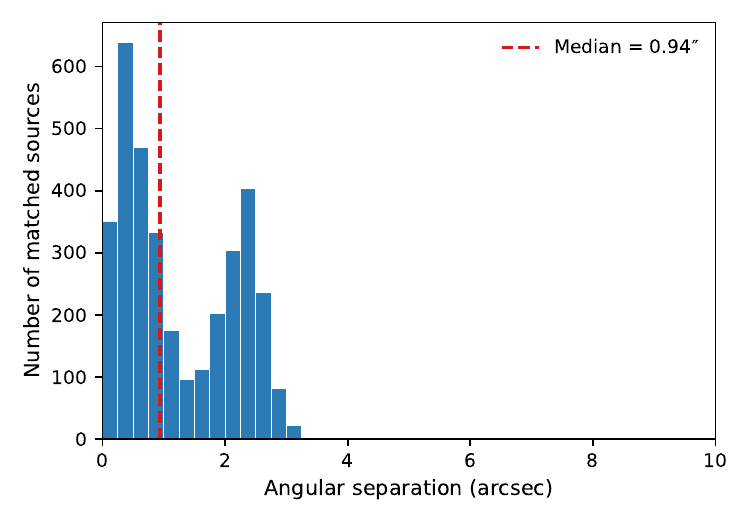}
\caption{Distribution of angular separations between independently
detected Vela sources and their nearest \citet{solano2022}
counterpart. The median separation of 0.94~arcsec reflects the
combined astrometric precision of both catalogs.
\label{fig:separation}}
\end{figure}

We cross-match our filtered 2{,}852{,}431-source candidate-vanished
catalog with the public 5{,}399-source unidentified-transient subset
reported by \citet{solano2022}, itself derived from a broader
298{,}165-source POSS-I-only workflow sample.
At a matching radius of 10~arcsec, we independently recover 3{,}450
sources (63.9\%), with a median angular separation of 0.94~arcsec
(Figure~\ref{fig:separation}). The separation distribution is sharply
peaked: 98.7\% of matches fall within 3~arcsec, indicating that
positional accuracy is not a limiting factor.

Of the 1{,}949 unrecovered Solano sources, 66 (3.4\% of unrecovered)
lie outside our survey footprint. For the remaining 1{,}883 within
our coverage, we classify each source by re-examining the raw
6.6-million-source pre-filter catalog:

\begin{itemize}
\item Approximately 54\% (${\sim}$1{,}010) are detected on the plate
in the raw catalog but rejected during PSF morphology
filtering---they appear as real sources but do not meet our
point-source criteria for FWHM or ellipticity.
\item Approximately 46\% (${\sim}$870) show no source-like detection
above our $5\sigma$ threshold at the cataloged position, suggesting
either that these sources fall below our detection limit or that
they are catalog-level artifacts.
\end{itemize}

Direct verification of all unrecovered Solano sources within our
coverage against Pan-STARRS~DR1 finds no counterpart within
3~arcsec. This result is consistent with the archive-absence
classification in \citet{solano2022}, but it should be interpreted as
a Pan-STARRS-specific validation rather than proof that the sources
are absent from every modern archive.

\begin{figure}
\centering
\includegraphics[width=\columnwidth]{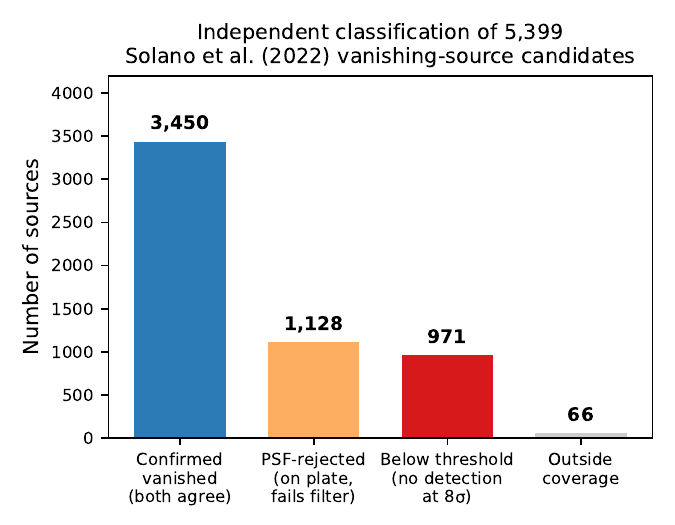}
\caption{Classification of all 5{,}399 \citet{solano2022} vanishing-source
candidates by independent detection status. 3{,}450 are independently
recovered; ${\sim}$1{,}010 of the unrecovered sources are detected in
our raw catalog but rejected by PSF filtering; ${\sim}$870 show no
detection above $5\sigma$; 66 fall outside survey coverage. None of
the unrecovered sources have a Pan-STARRS~DR1 counterpart within
3~arcsec.
\label{fig:breakdown}}
\end{figure}

\subsection{Full-Sky Survey}
\label{sec:fullsky}

The full-sky sweep was executed on four rented A100-equipped nodes, each
scanning a quarter of the RA range ($0\degr$--$360\degr$) at
Dec~$-33\degr$ to $+90\degr$. For each of the 122{,}991 grid
positions, the pipeline downloaded three 30~arcmin FITS cutouts from
the STScI Digitized Sky Survey---POSS-I Red, POSS-I Blue, and
POSS-II Red---detected sources on each cutout, applied PSF morphology
filtering, computed a local affine registration correction between
epochs, and flagged sources present on POSS-I Red with no counterpart
within 8~arcsec on \emph{either} reference epoch as transient
candidates.

The sweep was executed with preliminary detection parameters
($5\sigma$ threshold, FWHM range $0.5$--$2.0 \times$ median,
ellipticity $< 0.5$) prior to the benchmark calibration described in
Section~\ref{sec:benchmark}. In benchmark testing these looser settings
produce substantially more false candidates than the calibrated
configuration, especially on same-plate control fields.
Across 932 plates spanning 380 unique observation dates (November
1949 to December 1958), the sweep yielded 6{,}646{,}874 transient
candidates---sources on POSS-I Red absent from both POSS-I Blue and
POSS-II Red at the preliminary threshold.

These candidates were then post-processed with stricter fixed
morphology cuts (catalog-measured FWHM 2--8 and ellipticity $< 0.3$),
deduplicated on a 3~arcsec grid between adjacent patches, and bulk
cross-matched against Pan-STARRS~DR1 via CDS~XMatch. Sources with a
modern counterpart within 3~arcsec were rejected. After this
filtering, 2{,}852{,}431 candidate vanished sources remain. Of the 380
observation dates, 368 fall within the 1949--1957 study window used
for temporal analysis.
Total compute cost for the sweep was approximately \$50~USD.

A full 30~arcmin reprocessing with the calibrated $8\sigma$ parameters
has not yet been performed. On benchmark fields, the calibrated settings
materially improve negative-control performance relative to the
preliminary settings, so the reprocessed catalog is expected to be
substantially smaller and cleaner. All results in
Sections~\ref{sec:solano}--\ref{sec:temporal} use the preliminary
catalog described above.

\subsection{Temporal Correlation with Nuclear Tests}
\label{sec:temporal}

Following the methodology of \citet{bruehl2025}, we test whether
the occurrence of candidate vanished sources is temporally associated with
above-ground nuclear weapons tests conducted during the POSS-I
survey period. \citet{solano2022} report a broader 298{,}165-source
POSS-I-only workflow sample, of which 5{,}399 unidentified transients
were released in the public archive. As discussed by
\citet{villarroel2026response}, the later ensemble temporal analyses in
\citet{bruehl2025} use an improved ${\sim}$107{,}000-source transient
sample derived from that broader workflow; the transient dataset
described in \citet{bruehl2025} contains 107{,}875 objects with
observation dates and times. This is the relevant published comparison
sample for temporal inference, rather than the public 5{,}399-source
subset used for our cross-match validation in Section~\ref{sec:solano}. Our temporal analysis uses the full
2.85-million-source catalog produced by our pipeline. We identify 126 above-ground tests
between November 1949 and April 1957 from the Johnston Archive. For calendar-day
summaries, we construct day sets relative to each test date (e.g.,
$[+1]$, $[-1,0]$); these sets are descriptive and can overlap when
tests cluster in time. For count modeling, we analyze the 368 nights
with POSS-I observations in this study interval.

\subsubsection{Calendar-Day Binary Analysis}

For each calendar day, we record whether any vanished source was
observed on that date. Of the 2{,}718 days, 368 (13.5\%) have at
least one vanished source. The baseline rate---days more than 3
days from any test---is 13.3\% (271/2{,}038 days). Before-test
windows show rates close to the baseline: 13.3\% at $[-2, -1]$~days
and 14.1\% at $[-1, 0]$~days. On the day immediately following a test
($[+1]$~day), the rate rises to 17.9\% (22/123 days), corresponding
to RR~$= 1.35$ with 95\% CI [0.91, 2.00]. Relative to the baseline,
the two-sided Fisher exact $p$-value is 0.17; under the published
directional hypothesis of \citet{bruehl2025}, the one-sided
$p$-value is 0.098. The rate at $[+1, +2]$~days is 15.4\%.

\begin{figure*}
\centering
\includegraphics[width=\textwidth]{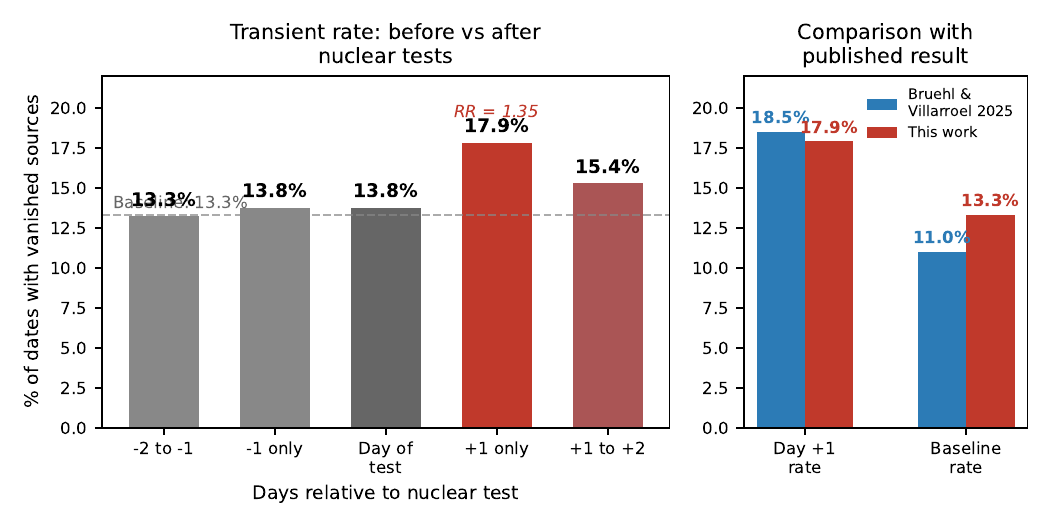}
\caption{Left: Bruehl-style calendar-day summary of study days with at
least one candidate vanished source, binned by relation to nuclear test
dates. The $[+1]$ window is elevated relative to the baseline
(17.9\% vs 13.3\%; RR~$= 1.35$), but the effect is not statistically
significant and is schedule-sensitive because all 368 study-window
observation nights contain at least one candidate. Right: published
\citet{bruehl2025} calendar-day comparison shown for context only.
\label{fig:temporal}}
\end{figure*}

This directional pattern---flat before tests, higher after---is
qualitatively similar to the result reported by \citet{bruehl2025},
but it should be interpreted cautiously here. All 368 study-window
observation nights contain at least one candidate vanished source
after Pan-STARRS filtering, so the binary calendar-day indicator
collapses to whether the survey observed that date. We therefore
report the $[+1]$ relative risk descriptively for comparison with
\citet{bruehl2025}, not as independent inferential evidence for a
post-test transient excess. The window comparisons in
Figure~\ref{fig:temporal} are exploratory and uncorrected for
multiple testing.

\subsubsection{Negative Binomial GLM}

To test whether the nuclear test window predicts the
\emph{number} of candidate vanished-source detections per observation night
(rather than simply their presence or absence), we fit a
negative binomial GLM to the 368 observation nights, following
the count-based approach of \citet{bruehl2025}. The response
variable is the number of candidate vanished-source detections on each
night. The predictor is a binary indicator for whether the
observation date falls within $\pm 1$~day of a nuclear test.
We include the nightly number of observed grid patches, computed from
the full patch summary, as an exposure term to account for the varying
sky coverage per night (range: 11--1{,}000; median: 278); for
\texttt{NegativeBinomialP}, the exposure is supplied on the
original scale and logged internally by the model.

Nightly counts are highly overdispersed relative to Poisson
(variance-to-mean ratio $\approx 4.1 \times 10^3$), supporting
the negative binomial specification. For the symmetric
$\pm 1$~day window, we find an incidence rate ratio
(IRR) of 1.03 (95\% CI: [0.89, 1.18]; $p = 0.71$).
Asymmetric windows are likewise null: the $[0, +1]$~day
window gives IRR~$= 1.01$ ($p = 0.90$), the $[-1, 0]$~day
window gives IRR~$= 0.98$ ($p = 0.82$), the
$[+1]$-only window gives IRR~$= 1.06$ ($p = 0.60$),
and the $[-1]$-only window gives IRR~$= 1.01$
($p = 0.95$). The fitted dispersion parameter for the
primary $\pm 1$~day model is $\alpha = 0.23$. None of the
five exploratory windows in Table~\ref{tab:nb_glm} reach
significance.

\begin{deluxetable}{lcccc}
\tablecaption{Negative binomial GLM results for nightly candidate
counts as a function of nuclear test proximity. Exposure term:
nightly patch coverage from the full patch summary
($N_{\mathrm{patches}}$, logged internally by the model).
$N_{\mathrm{in}}$ is the number of observation nights in the
specified window; $p$-values are two-sided and uncorrected across
the five exploratory windows. \label{tab:nb_glm}}
\tablehead{
\colhead{Window} & \colhead{$N_{\mathrm{in}}$} &
\colhead{IRR} & \colhead{95\% CI} & \colhead{$p$}
}
\startdata
$\pm 1$~day   & 52 & 1.03 & [0.89, 1.18] & 0.71 \\
$[0, +1]$     & 38 & 1.01 & [0.86, 1.19] & 0.90 \\
$[-1, 0]$     & 34 & 0.98 & [0.83, 1.16] & 0.82 \\
$[+1]$ only   & 22 & 1.06 & [0.86, 1.30] & 0.60 \\
$[-1]$ only   & 17 & 1.01 & [0.80, 1.27] & 0.95 \\
\enddata
\end{deluxetable}

The calendar-day summary and the nightly-count model answer different
questions, but in the present dataset the former is dominated by the
observation schedule. Because every study-window observation night
contains at least one candidate vanished source, the binary day-level
outcome cannot distinguish ``night with observations but no
candidates'' from ``night not observed.'' The negative binomial model
therefore provides the more relevant inferential check here, and it is
null across all tested windows.

We retain the temporal section because it identifies a concrete
limitation of the current full-sky catalog. The 1949--1957 nightly
counts average ${\sim}$7{,}250 candidate vanished sources, far above
the benchmark false-positive floor. As noted in
Section~\ref{sec:fullsky}, the full-sky sweep used preliminary
detection parameters; a reprocessing with the calibrated $8\sigma$
settings should substantially lower the candidate baseline. The
present null result establishes that any temporal claim depends
sensitively on filtering methodology and explicit control of the
observing schedule.

\section{Discussion}
\label{sec:discussion}

The 63.9\% independent recovery rate against \citet{solano2022}
shows that our automated source-level analysis reproduces most
catalog-level vanishing-source detections despite using a
fundamentally different pipeline. The 0.94~arcsec median positional
agreement also argues against the matches being chance coincidences.

The 36.1\% of Solano sources not recovered by our pipeline divide into
two informative categories after excluding the 66 sources outside our
footprint. Roughly half of the remaining unrecovered sources are
present in our raw catalog but rejected by the PSF filter, indicating a
methodological disagreement about source morphology. Some may be
genuine astrophysical transients with non-stellar morphology (e.g.,
resolved sources or overlapping objects); others may reflect
contamination or thresholding differences between the two pipelines.
The other roughly half show no raw detection above our $5\sigma$
threshold and require further investigation to determine whether they
represent very faint real sources below our detection limit or
catalog-level artifacts in the original survey.

The 0\% Pan-STARRS match rate among unrecovered Solano sources is a
notable positive result for the VASCO program: whatever these sources
are, they are not explained by an obvious Pan-STARRS counterpart. This
supports the narrower claim that many catalog entries lack a modern
optical match at Pan-STARRS depth; it does not, by itself, establish
astrophysical disappearance.

The temporal correlation with nuclear weapons testing
(Figure~\ref{fig:temporal}) yields a narrower conclusion than the
current draft implied. The Bruehl-style calendar-day comparison shows
a post-test elevation at $[+1]$ day (RR~$= 1.35$). But the confidence
interval is wide and includes unity, and the binary indicator is
schedule-sensitive because all 368 study-window observation nights
contain at least one candidate vanished source. In our data, the
statistic is therefore equivalent to asking whether POSS-I happened to
observe on that date.

The correctly parameterized negative binomial model, which uses nightly
candidate counts with sky coverage as an exposure term, is null for
the primary $\pm 1$ window and for all asymmetric windows in
Table~\ref{tab:nb_glm}. These null results do not rule out a weaker
association in a cleaner catalog, but they do show that the present
data do not independently support a count-level temporal effect. Any
future reanalysis should use calibrated filtering, explicit control
for the observation schedule and plate-level clustering, and
pre-specified windows to avoid over-interpreting exploratory
comparisons.

\section{Conclusions}
\label{sec:conclusions}

We have demonstrated that a fully automated pipeline can independently
recover 63.9\% of published POSS-I vanishing-source candidates with
sub-arcsecond positional accuracy, while maintaining low false positive
rates on benchmark fields. The pipeline recovers the well-studied
April 1950 and July 1952 transient fields and confirms that unmatched
catalog sources in our footprint generally lack obvious
Pan-STARRS counterparts.

For the temporal association with nuclear weapons tests, our
independently constructed 2{,}852{,}431-source catalog yields only
inconclusive evidence. The directly comparable published benchmark for
this ensemble test is the ${\sim}$107{,}000-source VASCO statistical
sample used by \citet{bruehl2025}, not the public 5{,}399-source subset
that we use above for catalog-overlap validation.
A Bruehl-style calendar-day comparison gives a descriptive post-test
asymmetry (RR~$= 1.35$, 95\% CI [0.91, 2.00]), but that statistic is
tied to the survey schedule in our dataset. A negative binomial model
of nightly counts with nightly patch coverage as exposure is null
(IRR~$= 1.03$, 95\% CI [0.89, 1.18], $p = 0.71$). The main robust
result of this study is therefore the independent catalog-level
replication; the nuclear-test question remains open and will require a
cleaner full-sky reprocessing plus explicit control of observational
confounds.

\begin{acknowledgments}
This work uses data from the STScI Digitized Sky Survey, the
Pan-STARRS1 survey, and the Spanish Virtual Observatory vanishing
sources catalog. We thank Enrique Solano for providing catalog access.
\end{acknowledgments}

\bibliography{references}

@article{bruehl2025,
  author  = {Bruehl, Jesse and Villarroel, Beatriz},
  title   = {A statistical analysis of the temporal and spatial distributions of {POSS-I} transient candidates},
  journal = {Scientific Reports},
  volume  = {15},
  pages   = {34125},
  year    = {2025},
  doi     = {10.1038/s41598-025-21620-3}
}

@article{villarroel2021,
  author  = {Villarroel, Beatriz and Imaz, Inigo and Joshi, Jayanant and others},
  title   = {Exploring nine simultaneously occurring transients on 12 {April} 1950},
  journal = {Scientific Reports},
  volume  = {11},
  pages   = {12794},
  year    = {2021},
  doi     = {10.1038/s41598-021-92162-7}
}

@article{solano2022,
  author  = {Solano, Enrique and Garc{\'{\i}}a-Alvarez, David and Marcos-Arenal, Pablo and others},
  title   = {Identification of vanishing objects in multi-epoch surveys},
  journal = {Monthly Notices of the Royal Astronomical Society},
  volume  = {515},
  pages   = {1380--1395},
  year    = {2022},
  doi     = {10.1093/mnras/stac1880}
}

@article{solano2024,
  author  = {Solano, Enrique and others},
  title   = {A bright triple transient that appeared on a {POSS-I} plate in {July} 1952},
  journal = {Monthly Notices of the Royal Astronomical Society},
  volume  = {527},
  pages   = {6312--6324},
  year    = {2024},
  doi     = {10.1093/mnras/stad3726}
}

@article{villarroel2020,
  author  = {Villarroel, Beatriz and Solano, Enrique and Mattsson, Lars and others},
  title   = {Our {Sky} Now and Then: Searches for Lost Stars and Impossible Effects as a Tool for Discovering New Astrophysics},
  journal = {The Astronomical Journal},
  volume  = {159},
  pages   = {8},
  year    = {2020},
  doi     = {10.3847/1538-3881/ab570f}
}

@article{lasker2008,
  author  = {Lasker, Barry M. and Lattanzi, Mario G. and McLean, Brian J. and others},
  title   = {The Second-Generation Guide Star Catalog: Description and Properties},
  journal = {The Astronomical Journal},
  volume  = {136},
  pages   = {735--766},
  year    = {2008},
  doi     = {10.1088/0004-6256/136/2/735}
}

@article{chambers2016,
  author  = {Chambers, K. C. and Magnier, E. A. and Metcalfe, N. and others},
  title   = {The {Pan-STARRS1} Surveys},
  journal = {arXiv e-prints},
  year    = {2016},
  eprint  = {1612.05560}
}

@article{villarroel2025profiles,
  author  = {Villarroel, Beatriz and Solano, Enrique and Marcy, Geoffrey W.},
  title   = {On the Image Profiles of Transients in the {Palomar Sky Survey}},
  journal = {arXiv e-prints},
  year    = {2025},
  eprint  = {2507.15896}
}

@article{villarroel2026response,
  author  = {Villarroel, Beatriz and Streblyanska, Alina and Bruehl, Stephen and Geier, Stefan},
  title   = {A Response to paper Critical Evaluation of Studies Alleging Evidence for Technosignatures in the {POSS1-E} Photographic Plates by Watters et al. (2026)},
  journal = {arXiv e-prints},
  year    = {2026},
  eprint  = {2602.15171}
}

\end{document}